\documentclass[10pt,journal]{IEEEtran}

\usepackage{graphicx}
\usepackage{amsmath}
\usepackage{amssymb}
\usepackage{float}
\usepackage{cite}
\usepackage{algorithm}
\usepackage{bibentry}
\usepackage{balance}
\usepackage{algorithm}
\usepackage{algorithmicx}
\usepackage{algpseudocode}
\usepackage{xcolor}
\usepackage{subfig}

\usepackage{url}

\graphicspath{{img/}}

\allowdisplaybreaks

\begin{document}

\bstctlcite{IEEEexample:BSTcontrol}

\title{LTE I/Q Data Set for UAV Propagation Modeling, Communication, and  Navigation Research \thanks{Sung Joon Maeng, Ozgur Ozdemir, \.{I}smail G\"{u}ven\c{c}, Mihail L. Sichitiu, Magreth Mushi, and Rudra Dutta are with North Carolina State University.}}

\author{\IEEEauthorblockN{Sung Joon Maeng, Ozgur Ozdemir, \.{I}smail G\"{u}ven\c{c}, Mihail L. Sichitiu, Magreth J. Mushi, and Rudra Dutta}}

\maketitle

\begin{abstract}
Recently, unmanned aerial vehicles (UAVs) have been receiving significant attention due to their wide range of potential application areas. To support UAV use cases with beyond visual line of sight (BVLOS) and autonomous flights,  cellular networks can serve as ground connectivity points, and they can provide remote control and payload communication for UAV links. However, there are limited data sets to study the coverage of cellular technologies for UAV flights at different altitudes and develop machine learning (ML) techniques for improving UAV communication and navigation. In this article, we present raw LTE I/Q sample data sets from physical field experiments in the Lake Wheeler farm area of the NSF AERPAW experimentation platform. We fly a UAV that carries a software-defined radio (SDR) at altitudes ranging from 30~m to 110~m and collect raw I/Q samples from an SDR-based LTE base station on the ground operating at 3.51 GHz. We adopt a standard metadata format for reproducing the results from the collected data sets. The post-processing of raw I/Q samples using MATLAB's 4G LTE toolbox is described and various representative results are provided. In the end, we discuss the possible ways that our provided data set, post-processing sample code, and sample experiment code for collecting I/Q measurements and vehicle control can be used by other ML researchers in the future. 
\end{abstract}

\section{Introduction}\label{sec:intro}
As the demand for unmanned aerial vehicles (UAVs) rapidly increases with their various use cases, cellular-connected UAVs are expected to be used widely in the future.  Due to their flexibility, cost-efficiency, and accessibility, UAVs can be used in a wide range of applications, from delivery to search and rescue, infrastructure inspections, wireless access points, among others~\cite{7463007}. To support such broad potential applications of UAVs, the use of long-range wireless communication systems such as cellular network technologies is critical.

To realize pervasive wireless connectivity to UAVs in the future, it is important to understand and model air-to-ground channel propagation realistically in practical conditions. In the literature, air-to-ground channel characteristics are typically measured using a UAV-mounted LTE cellular smartphone, and smartphone software such as Keysight NEMO Outdoor, Rohde \& Schwarz QualiPoc, and InfoVista TEMS is utilized to collect data sets~\cite{8746290,8108204,8369158}. However, such commercial smartphone software provides only a limited set of key parameters of interest (KPIs), they are costly to acquire, and they do not provide access to raw I/Q data samples. There are also various works that study machine learning (ML) techinques for UAV navigation based on signal strength observations (see e.g.~\cite{chowdhury20203,9354009}); however, the RF data are typically based on simulations and not real-world measurements. 

In this article, we provide a software-defined-radio (SDR) based framework for collecting and post-processing I/Q samples from an LTE eNB. The receiver SDR is carried at a programmable UAV. We carried out field experiments using the NSF aerial experimentation and research platform for advanced wireless (AERPAW) testbed, and the data is uploaded to IEEE Dataport~\cite{IEEEDataPort} for wider access and use by the research community. 
The use of raw I/Q samples gives the flexibility to obtain any KPIs of interest using software such as MATLAB's LTE Toolbox. We adopt the SigMF metadata format for sharing the I/Q data publicly~\cite{SigMF}. We describe the post-processing of the collected I/Q samples and produce representative results to explore air-to-ground channel propagation characteristics and different receiver algorithms. 

Our main contributions in this article are as follows:
\begin{itemize}
  \item[--] We collect raw LTE I/Q samples from UAV experiments at five different UAV altitudes in a rural environment and provide the collected data publicly for use by other experimenters. 
  \item[--] We format and present the collected I/Q sample and GPS data using the SigMF metadata standard format. 
  \item[--] We elaborate on the post-processing of the collected I/Q samples using MATLAB's LTE Toolbox and provide representative results.
  \item[--] We describe various different ways that the collected I/Q data, sample experiment codes, and post-processing codes can be used by other wireless and ML researchers working in related areas. 
\end{itemize}
Our earlier work~\cite{maeng2022aeriq} provides additional results and  discussions related to air to ground channel propagation and modeling.

\section{Experiment Description}\label{sec:}
In this section, we briefly overview the AERPAW platform, describe the I/Q sample collection experiment, and summarize the specific measurement scenario and the experiment setup.

\subsection{AERPAW Platform Description}
\begin{figure}
    \centering
    \subfloat[Lake Wheeler Road Field Labs site.]{\includegraphics[width=0.24\textwidth]{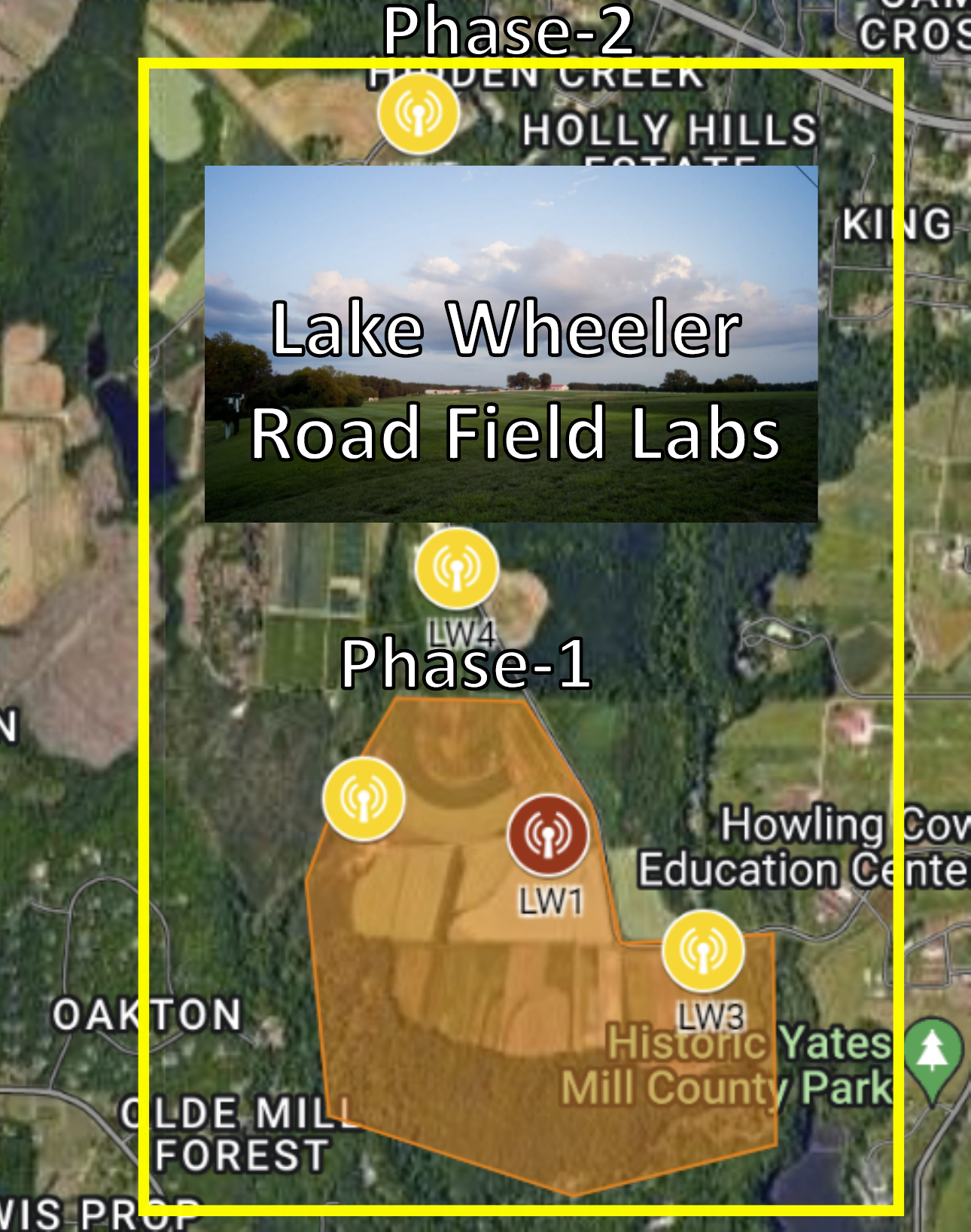}\label{fig:AERPAW_site_LW}}~
    \subfloat[Centenninal Campus site.]{\includegraphics[width=0.24\textwidth]{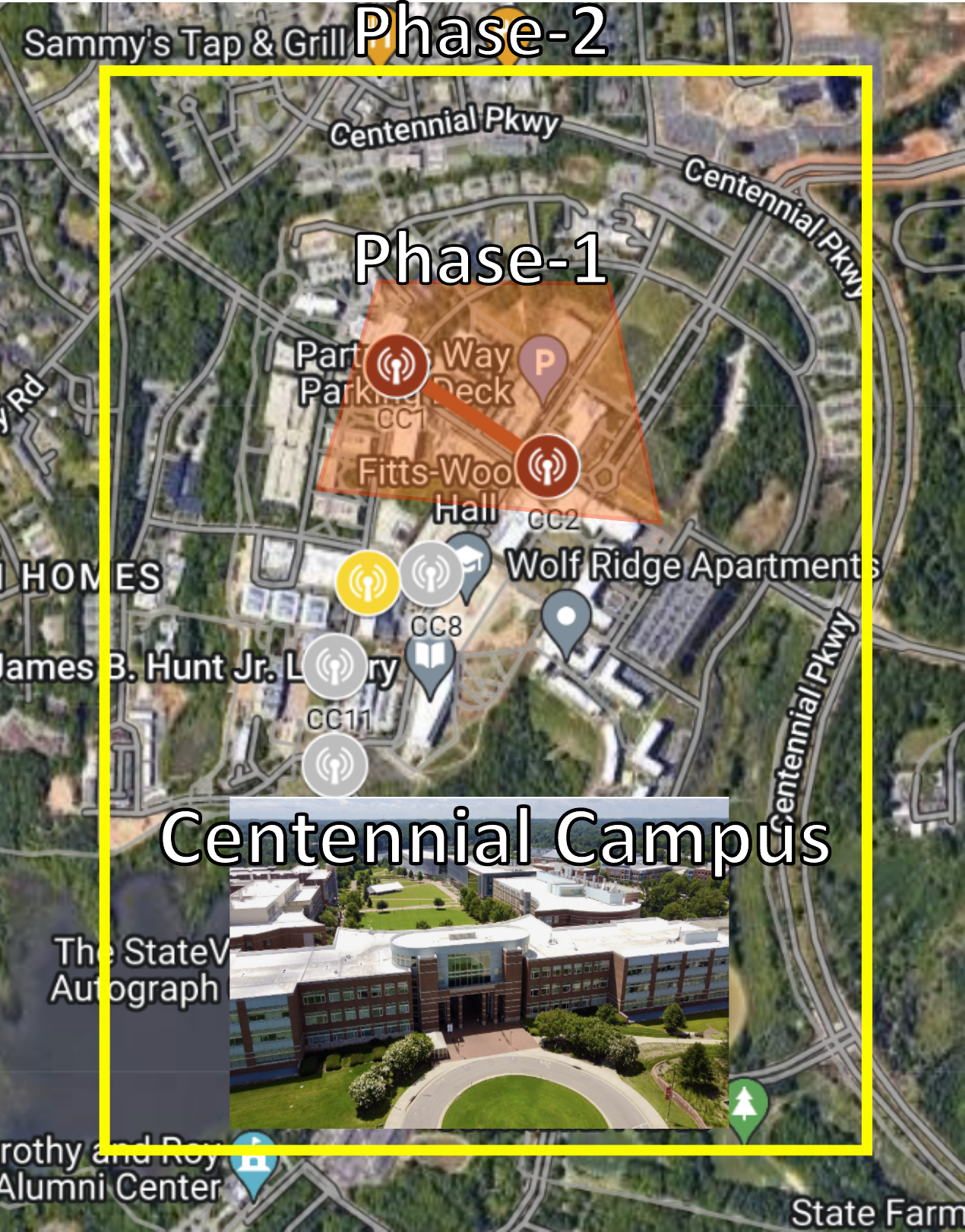}}
    
    \subfloat[UAV measurements at Lake Wheeler Road Field Labs]{\includegraphics[width=0.49\textwidth]{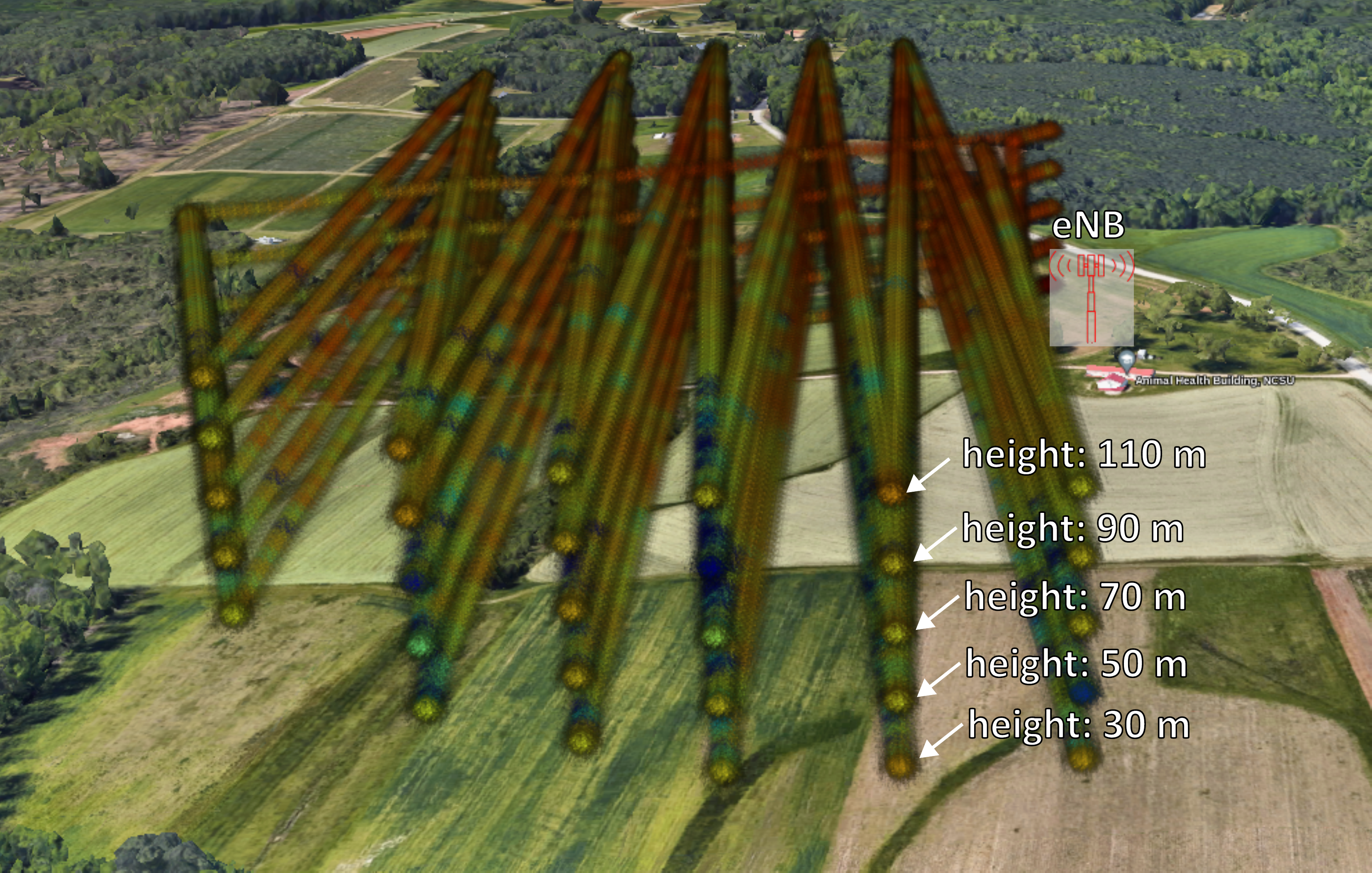}\label{fig:trajectory_Googlemap}}
    \caption{(a) AERPAW Lake Wheeler Road Field Lab outdoor deployment; (b) AERPAW Centennial Campus outdoor deployment; (c) Trajectories and reference signal received power (RSRP) for at UAV on  five different altitudes. Jet colormap is used to represent the strength of RSRP (red is the highest and  blue is the lowest RSRP).}
    \label{fig:AERPAW_site}
\end{figure}
\begin{figure}
    \centering
    \subfloat[LTE I/Q samples collection and post-processing scenario.]{\includegraphics[width=0.48\textwidth]{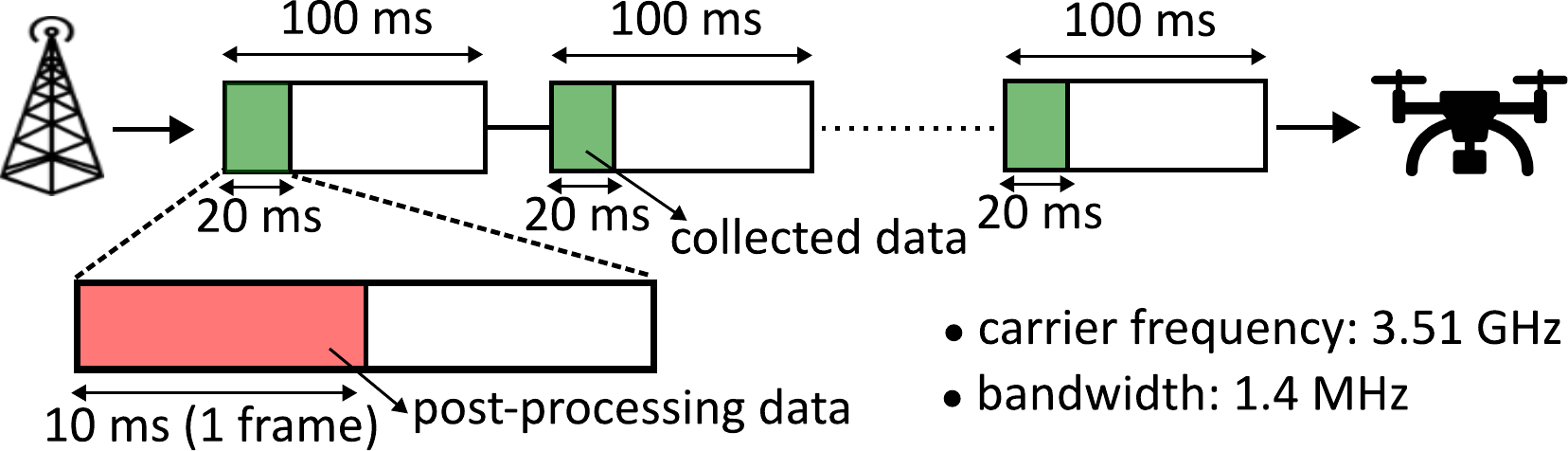}\label{fig:frame_structure}}
    
    \subfloat[Received signal power in LTE resource grid of 6~ms duration.]{\includegraphics[width=0.48\textwidth]{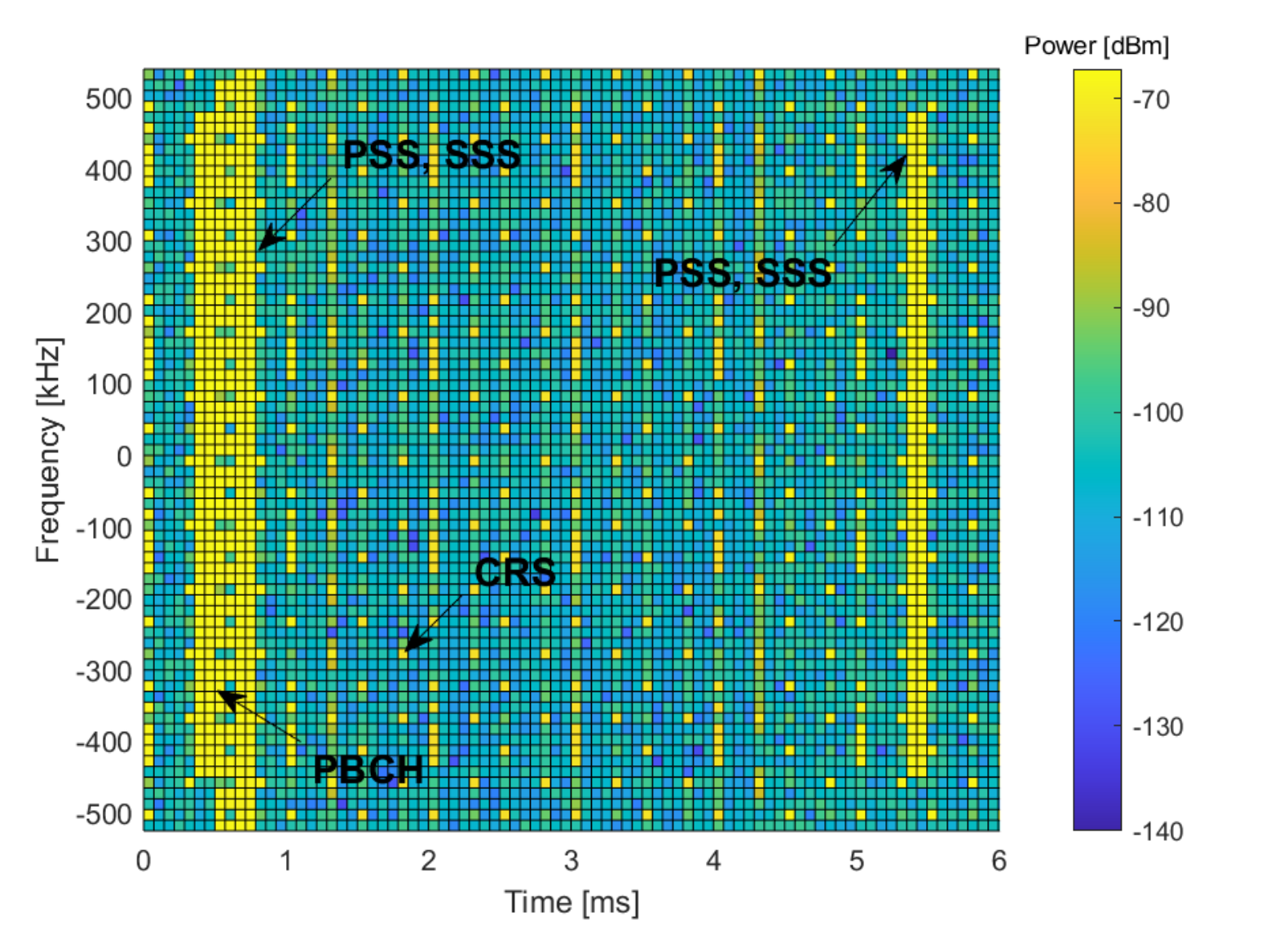}\label{fig:resource_grid}}
    \caption{(a) LTE I/Q data sets collection scenario, and (b) received signal power in LTE resource grid generated from raw I/Q samples using MATLAB's LTE Toolbox.}\label{fig:data_collection_scenario}
\end{figure}

AERPAW is an advanced wireless research testbed where experimenters are able to execute physical field experiments using programmable radios and vehicles. Experimenters can work on their experiments remotely in a development environment and once they are ready, the experiments are executed by the AERPAW operations team in the field. The collected measurement data are available to experimenters to view and analyze back in the virtual environment, e.g., comparing with the results obtained in the emulation environment. AERPAW offers various experiment scenarios and sample experiments for its users with fixed and portable nodes as well as autonomous UAVs~\cite{9438449}. In addition, AERPAW experiment sites cover rural (Lake Wheeler Road Field Labs), suburban, and urban (Centennial Campus) environments as shown in Fig.~\ref{fig:AERPAW_site}. The Phase-1 AERPAW fixed radio nodes (red circles) have been generally available as of November 2021, while the Phase-2 AERPAW fixed radio nodes (yellow circles) are expected to be generally available in early 2023.

AERPAW supports various types of universal software radio peripheral (USRP) SDR platforms from National Instruments (NI) for AERPAW fixed radio nodes. Moreover,  UAVs, helikites, and unmanned ground vehicles (UGVs) can be used to carry mobile portable nodes that can include an SDR, a mobile phone, and wireless sensors.

\subsection{I/Q Data Collection Sample Experiment}

The I/Q sample collection experiment of the AERPAW platform~\cite{AERPAW_site} considers a specified center frequency and bandwidth, and it records IQ samples (along with GPS coordinates, when applicable) at a USRP. The USRP can be at a fixed node or it can be carried by a UAV. The data is stored in MATLAB (.mat) file format and it can be post-processed by MATLAB's toolboxes or other relevant software. Generally speaking, the I/Q samples using the provided sample software~\cite{AERPAW_site} can be collected at multiple center frequencies with configured sampling rates for a specified amount of duration. An experimenter can configure USRP Rx gain, sampling rate, center frequencies, duration of I/Q waveform, wait time between each I/Q collection, and experiment duration. The user can  set up the experiment parameters in development mode. The dataset used for this work was obtained by using  an AERPAW fixed radio node configured as an LTE eNB and an AERPAW UAV-mounted portable node configured as an IQ sample collection device.

\subsection{Measurement Scenario and Setup}

\begin{figure*}
    \centering
    \subfloat[Steps from collecting RF and GPS data sets during an experiment, to publishing SigMF formatted data sets online.]{\includegraphics[width=0.75\textwidth]{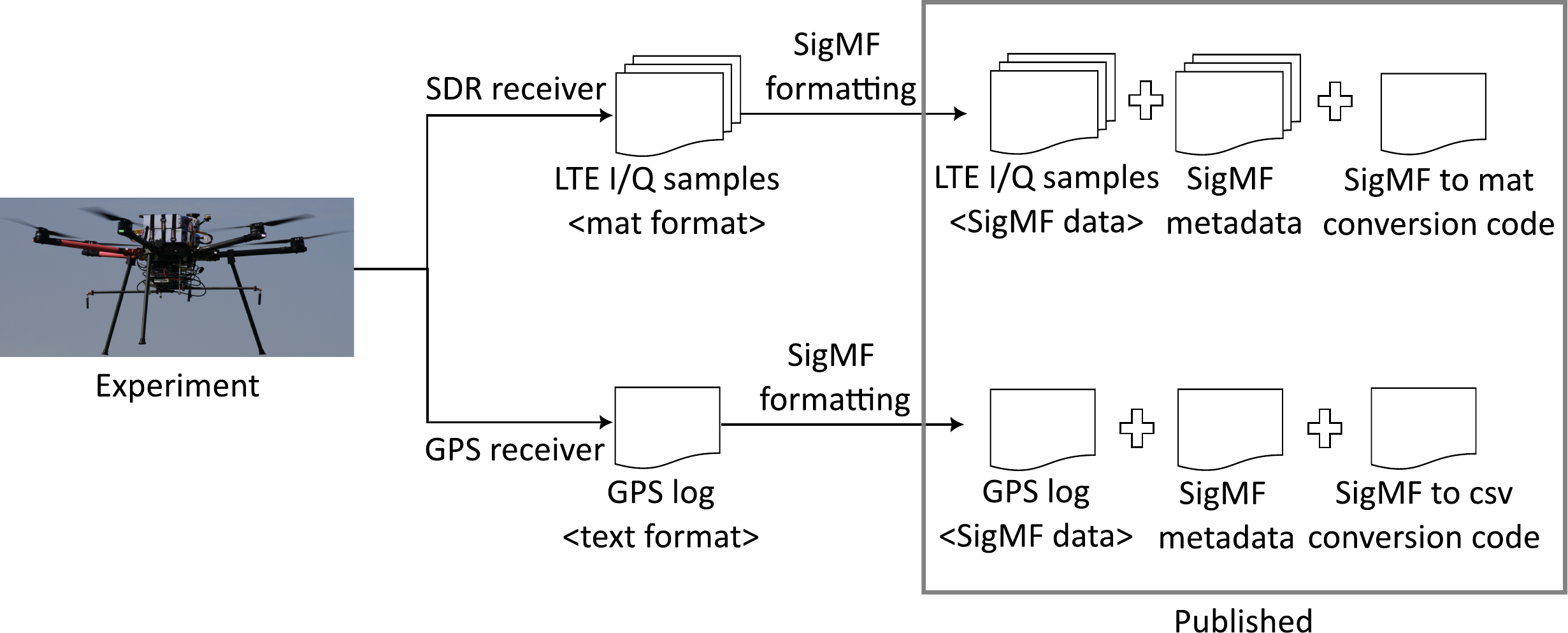}}

    \subfloat[Data structure of I/Q samples and GPS data, and conversion between mat and SigMF formats.]{\includegraphics[width=0.75\textwidth]{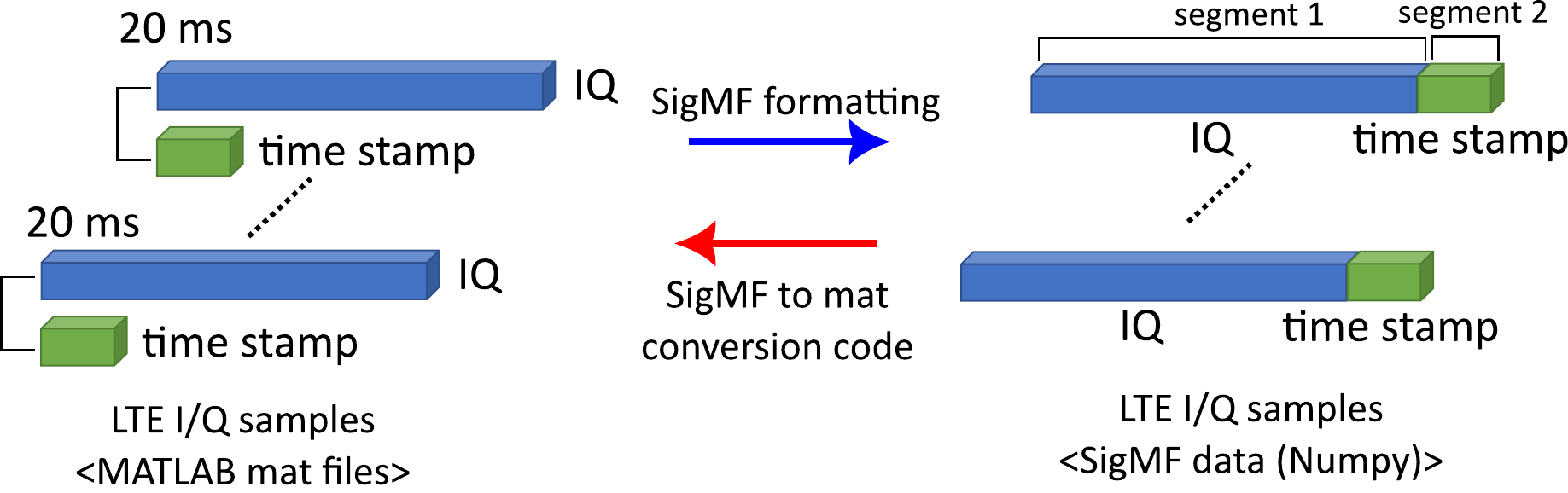}}
    \caption{(a) Procedure of data collection and SigMF data formatting; (b) Structure of the  data provided in this article. The conversion codes between SigMF and .mat formats are also provided as part of the IEEE Dataport submission.}
    \label{fig:SigMF_flow}
\end{figure*}

In this article, using the I/Q data collection sample experiment in~\cite{AERPAW_site} described earlier, we conduct measurements using a UAV with predefined trajectories at five different altitudes. We collect LTE I/Q samples at the AERPAW Lake Wheeler Road Field Labs site which can be classified as an open rural area. A fixed radio node at the Lake Wheeler Road Field Labs (LW1 colored red in Fig.~\ref{fig:AERPAW_site_LW}) is configured to be an LTE eNB and a UAV-mounted SDR and a GPS receiver is configured to be a portable node. The srsRAN open-source SDR software is used as an LTE eNB at the fixed radio node, and both the eNB and the portable node use USRP B205mini. The height of the eNB tower is 10 m and both the eNB and the UAV are equipped with a vertically oriented single dipole antenna. The center frequency is fixed at 3.51~GHz and the bandwidth of the transmitted signal is 1.4~MHz. During the experiments, there is no other signal source observed within this spectrum. 

The USRP at the fixed radio node acting as an eNB continuously transmits common reference symbols (CRSs), synchronization signals (SS), and physical broadcast channels (PBCH) during the experiment, while the UAV carrying the portable node equipped with a USRP running I/Q Data collection sample experiment collects 20~ms segments of data with 2 MHz sampling rate out of every 100 ms as described in Fig.~\ref{fig:data_collection_scenario}. The experiments are conducted multiple times with different altitudes (heights) of the UAV from 30~m to 110~m at increments of 20~m. In each flight, the UAV flies the identical predefined trajectory with a fixed height where the UAV sweeps the experiment site and flies back to the position where it takes off. After the UAV takes off, it flies south and north in a zigzag pattern. Then, it returns to the starting point. Fig.~\ref{fig:trajectory_Googlemap} shows the measured RSRP from five different experiments overlayed on the 3D view of Google Maps. In particular, the UAV flies at 30~m to 110~m altitudes in five separate missions, and the corresponding RSRP is obtained by post-processing the collected I/Q samples along with the GPS logs from the UAV. 

\begin{figure*}
    \centering
    \includegraphics[width=0.75\textwidth]{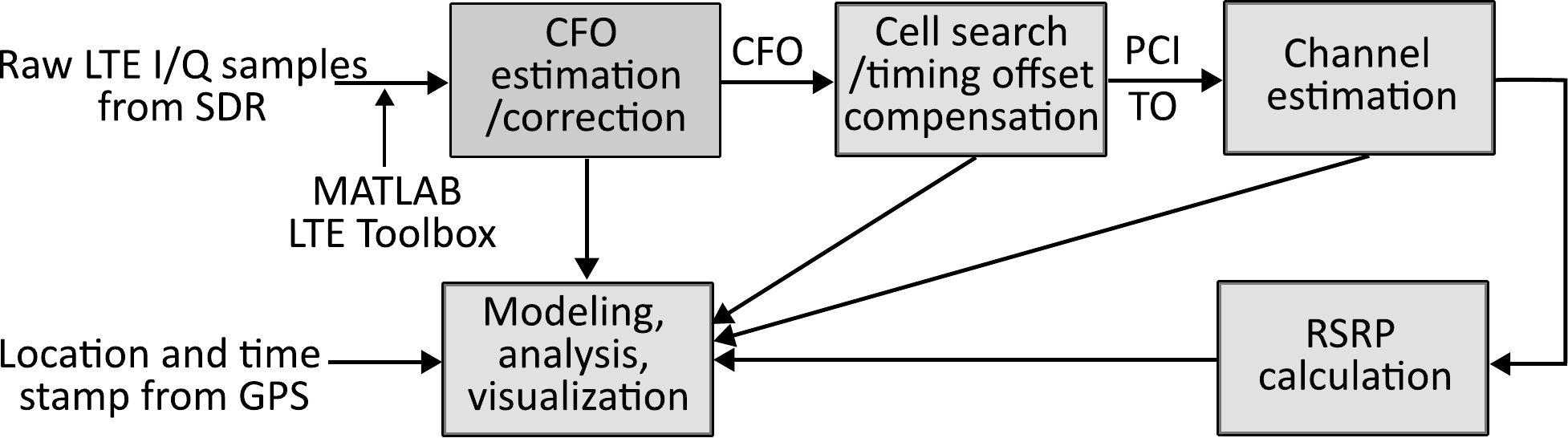}
    \caption{Block diagram for post-processing of I/Q samples and GPS log collected at an SDR and a GPS that are carried by a UAV.}
    \label{fig:post_processing_block}
\end{figure*}

\section{SigMF Data Formatting}\label{sec:}
In this section, we discuss how we format  the collected data sets using a standard data format and  share them publicly for use by other researchers. The Signal Metadata Format (SigMF) is a standard data format to describe the collected (recorded) data to those who would like to access them~\cite{SigMF}. SigMF prevents the loss of detailed information about the collected data over time and makes data sets compatible with different tools. SigMF recordings consist of any time-varying data source (e.g. collected I/Q samples) and a metadata file containing plain text that describes the data set.

\subsection{Converting I/Q samples and GPS Logs to SigMF}

In the experiments, the output data sets from the UAV portable node consist of 1) .mat files including LTE I/Q samples from the SDR receiver, and 2) a text file including GPS log from the GPS receiver. The .mat file for the LTE I/Q samples include complex numbers involving I/Q samples as well as the corresponding time stamps when those samples are collected. The GPS log provides the UAV's latitude, longitude, altitude, and time stamps during the experiments. To publish the measured data sets publicly for use by other researchers, we adopt the SigMF standard. We convert raw data sets to SigMF by using the SigMF Python package. Each I/Q .mat file and GPS log text file is rewritten by NumPy array SigMF data set and each generated SigMF data set is described by SigMF metadata. The collected I/Q samples and the GPS log in the SigMF data set are recorded in the complex-number floating-point 64-bits format and the real-number floating-point 64-bits format, respectively. 

The SigMF Metadata is written in JavaScript object notation (JSON) format and three JSON objects (global, capture, annotations) are used to describe the SigMF data set. The global object  provides information applicable to the entire data set such as the datatype of the SigMF data set, the version of SigMF, the author, and a general description of the data set. The segments of the data set can be described by capture and annotation objects. 

\subsection{Public Sharing of SigMF Formatted Data Sets}
After we obtain the SigMF format recordings (SigMF data set + SigMF metadata) from the original data sets, we can share them publicly so that one can download and use these data sets. However, most downloaders may need to use the original .mat file format with I/Q samples and read GPS logs in text file format. In particular, since the post-processing code is written based on the original data sets format, interested researchers who want to use the post-processing codes will need to obtain the data sets in their original format. Therefore, we also provide a SigMF to .mat format conversion Python code for I/Q samples data sets and a SigMF to comma-separated values (CSV) format conversion Python code for the GPS log. Fig.~\ref{fig:SigMF_flow} shows the flowchart from collecting data sets of an experiment to publishing the SigMF formatted data sets online, as well as the data structure of raw I/Q samples and SigMF formatted data sets.

\section{I/Q Data Post-Processing and Representative Results}\label{sec:}
In this section, we discuss the post-processing of the LTE I/Q samples and the GPS log obtained from the experiments, and we provide some representative results. 

\subsection{MATLAB LTE Toolbox for Post-Processing}
MATLAB LTE Toolbox~\cite{LTE_Toolbox} makes it possible to decode raw LTE I/Q samples from the experiments. In particular, we can synchronize with the LTE received signal, implement LTE cell search procedure, estimate the channel using CRS, and calculate RSRP from the raw I/Q samples. 
Furthermore, we can integrate the post-processed LTE signals into the GPS information to model and analyze the numerous characteristics of channel propagation, and visualize the received signal strength in the 3D map as shown in Fig.~\ref{fig:trajectory_Googlemap}. 

The block diagram for the post-processing of LTE I/Q samples is described in Fig.~\ref{fig:post_processing_block}. By MATLAB LTE Toolbox, we first estimate and compensate the carrier frequency offset (CFO) from input LTE I/Q samples. The cyclic prefix (CP) of the OFDM symbols is used in estimating the CFO. Then, we jointly estimate the physical layer ID and timing offset (TO) from a primary synchronization signal (PSS) by using the auto-correlation property of the synchronization signal and estimate the physical layer group ID from a secondary synchronization signal (SSS). By combining the physical layer ID and physical layer group ID, we can determine the physical cell identities (PCIs). By using the PCI, we can extract CRS corresponding to that PCI value, estimate the channel using the CRS, and calculate the RSRP. The detailed procedures of the post-processing can be found in \cite{maeng2022aeriq}. 

When we integrate raw I/Q samples with GPS information, we use the individual time stamps from the SDR and the GPS. Specifically, the SDR collects I/Q samples with time stamps, while the GPS records the locations with time stamps separately. Therefore, by matching and interpolating two different time stamps from the SDR and GPS receivers, we can integrate I/Q samples with GPS location information.

\subsection{Representative Results}
\begin{figure}
    \centering
    \subfloat[Path loss analysis from the RSRP.]{\includegraphics[width=0.48\textwidth]{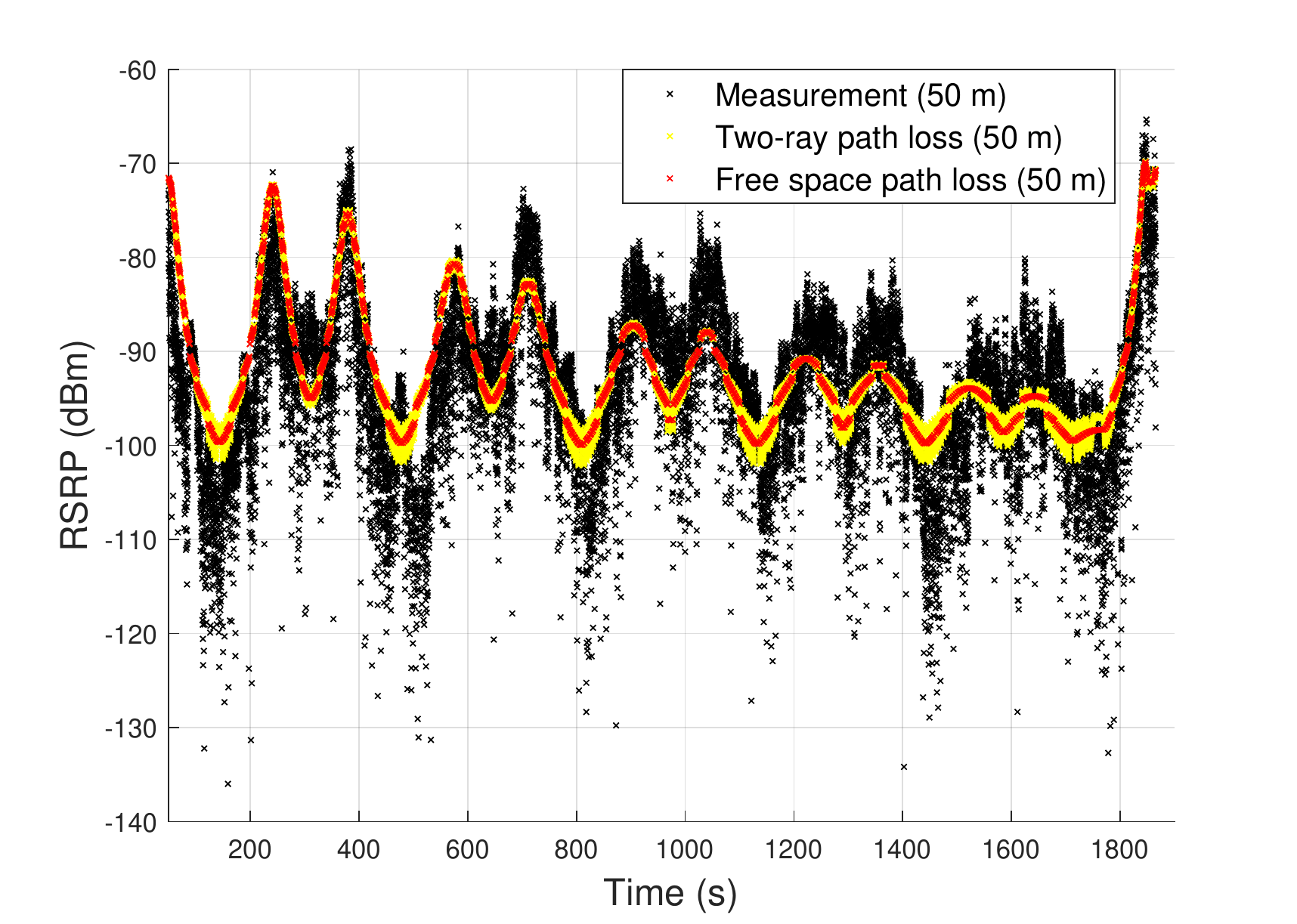}\label{fig:RSRP_PL}}

    \subfloat[Shadowing distribution analysis from the RSRP.]{\includegraphics[width=0.48\textwidth]{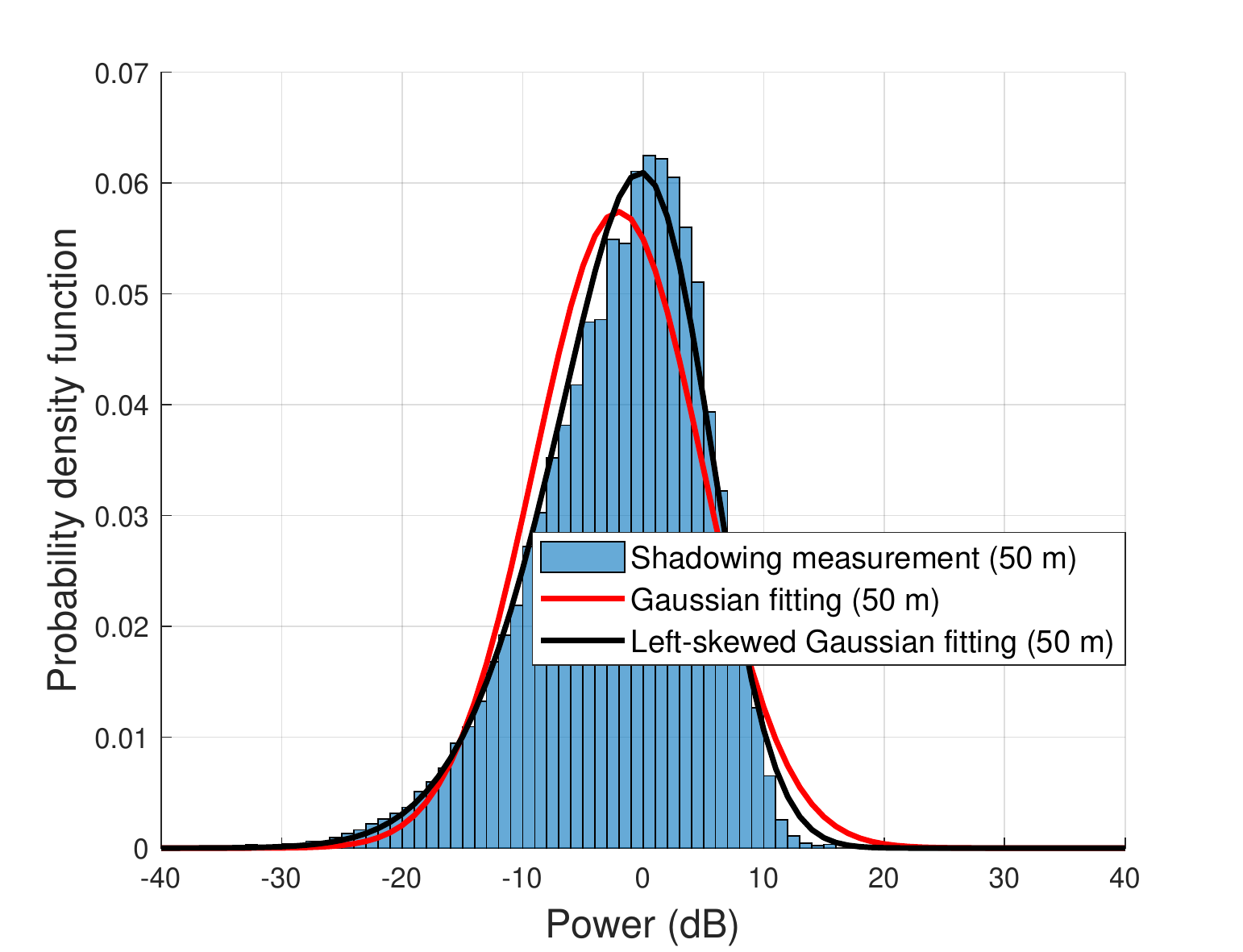}\label{fig:RSRP_shad}}
    \caption{Received signal (RSRP) analysis for 50~m UAV height experiment. The path loss and shadowing are fitted by models.}
    \label{fig:RSRP_results}
\end{figure}
\begin{figure}
    \centering
    \subfloat[90 m height at 380 s (RSRP high)]{\includegraphics[width=0.48\textwidth]{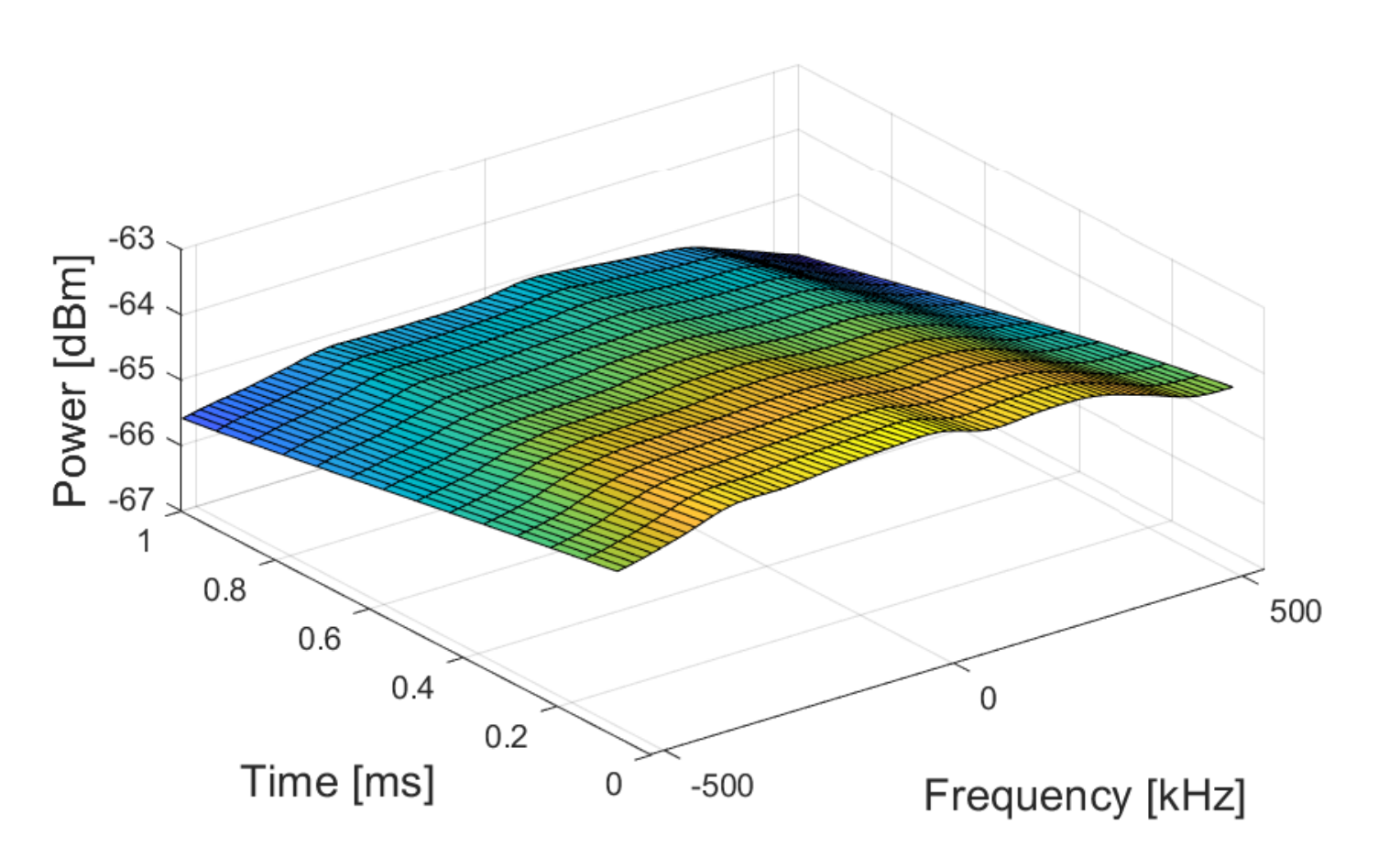}\label{fig:chan_est_high}}

    \subfloat[90 m height at 520 s (RSRP low)]{\includegraphics[width=0.48\textwidth]{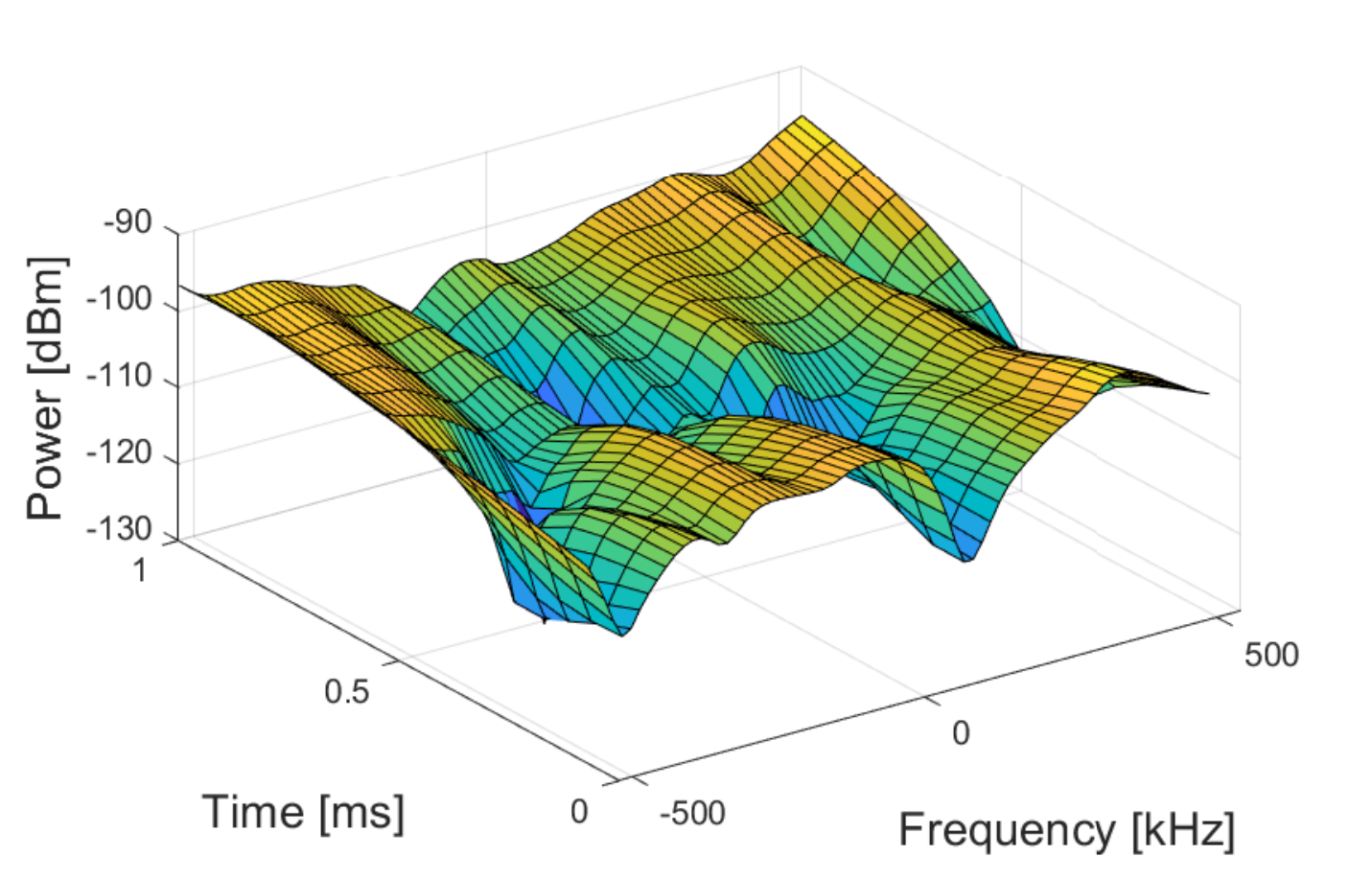}\label{fig:chan_est_low}}
    \caption{Estimated channel when RSRP is high (a) and low (b) from UAV height 90~m experiment. The CRS is used to estimate the channel and the unknown channels in time and frequency domains are interpolated.}
    \label{fig:chan_est}
\end{figure}
We show the key exemplary results of air-to-ground channel propagation analysis from the post-processing of LTE I/Q samples. Fig.~\ref{fig:RSRP_results} shows the analysis and fitting results of the received signal power (RSRP) from 50~m UAV height experimental data sets. The received signal power is modeled by the path loss component and shadowing component as follows:
\begin{align*}
    \mathsf{RSRP}&=\mathsf{P}_{\rm Tx}-\mathsf{PL}+w,
\end{align*}
where $\mathsf{P}_{\rm Tx}$, $\mathsf{PL}$, and $w$ represent transmit power, path loss, and shadowing. In Fig.~\ref{fig:RSRP_PL}, the measured RSRP is fitted to the free space path loss model and the two-ray path loss model. In both path loss models, we consider elevation angle-dependent Tx / Rx antenna radiation patterns. In the two-ray path loss model, a strong ground reflection path, as well as a line-of-sight (LoS) path, is considered in calculating received signal strength. Because of the ground reflection path, the fluctuation of the RSRP is captured by the two-ray path loss model compared with the free space path loss model. In Fig.~\ref{fig:RSRP_shad}, the measured shadowing component is fitted to Gaussian distribution and left-skewed Gaussian distribution. The measured shadowing component is calculated by
\begin{align*}
    \hat{w}&=\mathsf{RSRP} - \mathsf{P}_{\rm Tx} + \mathsf{PL}_{\rm 2ray},
\end{align*}
where $\mathsf{PL}_{\rm 2ray}$ denotes the two-ray path loss model which is obtained from Fig.~\ref{fig:RSRP_PL}. It is observed that although the measured shadowing is fitted well with the Gaussian distribution, the left-skewed Gaussian distribution fits the measured shadowing better than the Gaussian distribution.

Fig.~\ref{fig:chan_est} shows the estimated channels from a single set of measured raw LTE I/Q samples at a time when the RSRP is high (Fig.~\ref{fig:chan_est_high}) and low (Fig.~\ref{fig:chan_est_low}). These plots correspond to the time instants of 380~s and 520~s when the RSRP is strong and weak, respectively, in Fig.~\ref{fig:RSRP_PL}. The channel estimation step on the post-processing is described in Fig.~\ref{fig:post_processing_block}. The extracted CRSs are utilized to estimate the channel coefficients at the allocated resources and the unknown locations of the channel coefficients in the resource grid are interpolated by cubic interpolation. It is observed that the estimated channel at the high RSRP is flat, while the estimated channel at the low RSRP varies quickly in both time and frequency domains.

Aside from the presented results, we can generate other interesting results from the post-processing such as UAV speed changes, spatial correlation, coherence bandwidth, and correlation peaks from the PSS and SSS~\cite{maeng2022aeriq}.

\section{Possible Uses of Data and Future AERPAW Experiments}\label{sec:}
The provided measurement data, sample post-processing code, and the I/Q data collection sample experiment can be used by other researchers in various different ways.

\textbf{Propagation Modeling:} Different ML-based air-to-ground propagation models can be developed using the provided data -- see e.g.~\cite{duangsuwan2021path} and several other recent works in the literature. Such models can take into account (if known) or learn in real time (if not available) the 3D antenna pattern and orientation at the transmitter and the receiver, altitude of the UAV, height of the base station tower, communication frequency and bandwidth, speed and tilt angle of the UAV, propeller speed at the UAV (as they may affect Doppler features), among other features. Coherence bandwidth, coherence time, and spatial consistency can be characterized for various scenarios.

\textbf{New Receiver Algorithms:} The provided data can be used to develop new time and frequency domain synchronization algorithms and channel estimation techniques suitable for aerial communication scenarios. While Kriging was used in~\cite{maeng2022aeriq} for spatial interpolation, there are various other signal processing and ML techniques in the literature, as discussed e.g. in~\cite{6364111}, which can be further explored for improved accuracy.

\textbf{New UAV Trajectory Designs and Drone Corridors:} The 3D UAV coverage heatmap of the ground base station can be used to optimize UAV trajectory e.g. for maximizing the data rate for a given mission plan under realistic channel conditions. When only waypoints are defined, UAVs need to accelerate and deaccelerate to fly on linear paths between waypoints while satisfying the mission duration and RF performance constraints. The coverage map of a given environment can help in optimizing the UAV waypoints~\cite{chowdhury20203} under realistic UAV mission plans, or test the efficiency of ML-based on-the-fly waypoint selection algorithms~\cite{9354009}. The provided data can also be used to design and optimize 3D drone corridors to support UAV missions with good wireless coverage. 

\textbf{I/Q Data Collection Experiments:} The I/Q collection sample experiment provided in~\cite{AERPAW_site} can be tailored to collect data from an LTE network at different frequencies and with different UAV trajectories at multiple altitudes based on the specific needs of an experimenter. Other waveforms than LTE, such as 5G, WiFi, and LoRa, among other technologies, can be used as signal sources where I/Q data is to be collected. The AERPAW platform at NC State University is available for supporting such experiments remotely for external users after they are tested in AERPAW's emulation environment. Alternatively, experimenters can use the provided software to run experiments in their own environments of interest with or without UAVs. The collected data can be post-processed using MATLAB's various toolboxes -- an example code for LTE is provided for regenerating the results in this paper and in~\cite{IEEEDataPort}. A 5G I/Q data set and post-processing code example will be provided in the near future.

\textbf{Emulation Experiments:} Prior to running the experiments in an outdoor testbed, the same RF and vehicle software can be initially developed and executed in an emulation environment such as AERPAW's development environment~\cite{panicker2021aerpaw}. At the time of writing of this paper, the RF software emulation environment in AERPAW supports a simple two-path multipath channel, up to 100 MHz of bandwidth, different center frequencies, MIMO support for 2x2 spatial multiplexing, and different antenna radiation patterns. The vehicle emulation supports defining fixed waypoints for one or more UAVs and/or unmanned ground vehicles, or alternatively, the UAVs/UGVs can make autonomous decisions in real time for choosing subsequent waypoints based on RF signal observations. Once the RF and vehicle software is fully developed, they can be moved to AERPAW's testbed environment. Experimenters can compare their results from emulation and testbed, and if needed, can develop and test code further in the emulation and testbed environments.

\section{Concluding Remarks}\label{sec:conclusion}
In this article, we provide LTE I/Q data sets to support research for UAV channel propagation modeling, communications, and navigation research. We collect LTE I/Q samples by a UAV in an open rural environment and share the data sets using the metadata standard SigMF. MATLAB's LTE Toolbox is used for the post-processing of the collected I/Q samples. We provide representative results using the collected I/Q and GPS data -- as an example, we compare the received signal strength measurements using various different path loss and shadowing models. We also elaborate on various possible ML-based research ideas that can take advantage of the provided LTE I/Q data set, or alternatively, that can use AERPAW's emulation and testbed environments for running additional experiments using the provided sample experiment software.  

\section*{Acknowledgement}
This research is supported in part by the NSF award CNS-1939334 and its associated supplement for studying NRDZs.

\bibliographystyle{IEEEtran}
\bibliography{IEEEabrv,references}

\section*{Biographies}
\vskip -2\baselineskip plus -1fil
\begin{IEEEbiographynophoto} {Sung Joon Maeng} (smaeng@ncsu.edu) received his BS and MS degrees both in Electrical and Electronic Engineering from Chung-Ang University, South Korea, in 2015 and 2017, and received his Ph.D. degree in Electrical and Computer Engineering at North Carolina State University, USA, in 2022. He is now working as a Postdoctoral Scholar at North Carolina State University. His main research interests are cellular-connected UAV communications and mmWave and sub-THz communications.
\end{IEEEbiographynophoto}
\vskip -2\baselineskip plus -1fil
\begin{IEEEbiographynophoto}{Ozgur Ozdemir} (oozdemi@ncsu.edu) received the BS degree in Electrical and Electronics Engineering from Bogazici University, Istanbul, Turkey, in 1999 and the MS and Ph.D. degrees in Electrical Engineering from The University of Texas at Dallas, Richardson, TX, USA, in 2002 and 2007, respectively. From 2007 to 2016 he was an Assistant Professor at Fatih University, Turkey, and worked as a Postdoctoral Scholar at Qatar University for 3.5 years. He joined the Department of Electrical and Computer Engineering at North Carolina State University (NCSU) as a visiting research scholar in 2017. He is now serving as an Associate Research Professor at NCSU. His research interests include software-defined radios, channel sounding for mmWave systems, digital compensation of radio-frequency impairments, and opportunistic approaches in wireless systems.
\end{IEEEbiographynophoto}
\vskip -2\baselineskip plus -1fil
\begin{IEEEbiographynophoto}{Ismail Güvenc (F'21)} (iguvenc@ncsu.edu) received
the Ph.D. degree in electrical engineering from the
University of South Florida in 2006.
He was with the Mitsubishi Electric Research Labs
in 2005, with DOCOMO Innovations from 2006
to 2012, and with Florida International University,
from 2012 to 2016. From 2016 to 2020, he has been an
Associate Professor, and since 2020, he has been a Professor, with the Department of Electrical and Computer Engineering of North Carolina
State University. He has published more than 300
conference/journal articles and book chapters, and
several standardization contributions. He coauthored/co-edited four books and he is an inventor/co-inventor of
some 30 U.S. patents. His recent research interests include 5G wireless
systems, communications and networking with drones, and heterogeneous
wireless networks.
\end{IEEEbiographynophoto}
\vskip -2\baselineskip plus -1fil
\begin{IEEEbiographynophoto}{Mihail L. Sichitiu} (mlsichit@ncsu.edu) received his Ph.D. degree in Electrical Engineering from the University of Notre Dame in 2001. His current research interests include wireless networks and communications for UAVs. In these systems, he is studying problems related to localization, time synchronization, emulation, routing, fairness, and modeling. He is teaching wireless networking and UAV courses. He is a professor in the department of Electrical and Computer Engineering at NC State University.
\end{IEEEbiographynophoto}
\vskip -2\baselineskip plus -1fil
\begin{IEEEbiographynophoto}
{Magreth J. Mushi} (mjmushi@ncsu.edu) is currently a Research Associate at North Carolina State University (NCSU) Computer Science department. She joined the department in August 2021. Before joining the department, she was TERNET’s Chief Executive Officer and a Researcher at Open University of Tanzania.  She graduated her PhD in Network Security from NCSU in 2016. Her area of research interest is network security as well as network administration and management in Software Defined Networks.
\end{IEEEbiographynophoto}
\vskip -2\baselineskip plus -1fil
\begin{IEEEbiographynophoto}{Rudra Dutta} (rdutta@ncsu.edu) received his Ph.D. in Computer Science from North Carolina State University, Raleigh, USA, in 2001. From 1993 to 1997 he worked for IBM as a software developer and programmer in various networking related projects. He has been employed as faculty since 2001 in the department of Computer Science at the North Carolina State University, Raleigh, since 2013 as Professor, and since Fall, 2018, serving as Associate Department Head. His current research interests focus on design and performance optimization of large networking systems, Internet architecture, wireless networks, and network analytics. He is a senior member of IEEE and a distinguished member (distinguished engineer) of ACM.
\end{IEEEbiographynophoto}

\end{document}